\documentclass[letterpaper]{article}
\usepackage[preprint]{preprint}
\usepackage[hyphens]{url}
\usepackage{graphicx}
\usepackage[numbers,sort&compress]{natbib}
\usepackage{caption}
\usepackage{amsmath,amssymb}
\usepackage{booktabs}
\usepackage{xspace}
\newcommand{\ours}{PTA-IRT\xspace}

\title{Efficient SWE Agent Benchmarking via Trajectory-Aware Evaluation}
\author{
    Kefeng Duan\textsuperscript{\rm 1},
    Dewu Zheng\textsuperscript{\rm 1},
    Yanlin Wang\textsuperscript{\rm 1}\corresponding,\\
    Xiwen Wang\textsuperscript{\rm 1},
    Ensheng Shi\textsuperscript{\rm 2},
    Xilin Liu\textsuperscript{\rm 2},
    Yuchi Ma\textsuperscript{\rm 2},\\
    Jiachi Chen\textsuperscript{\rm 3},
    Mingwei Liu\textsuperscript{\rm 1},
    Zibin Zheng\textsuperscript{\rm 1}
}
\affiliations{
    \textsuperscript{\rm 1}School of Software Engineering, Sun Yat-sen University, Zhuhai, China\\
    \textsuperscript{\rm 2}Huawei Cloud Computing Technologies Co., Ltd., China\\
    \textsuperscript{\rm 3}College of Computer Science and Technology, Zhejiang University, Hangzhou, China\\
    \{duankf, zhengdw5, wangxw86\}@mail2.sysu.edu.cn,
    \{wangylin36, liumw26, zhzibin\}@mail.sysu.edu.cn\\
    \{shiensheng, liuxilin3, mayuchi1\}@huawei.com,
    chenjiachi@zju.edu.cn
}

\begin{document}

\pagestyle{plain}
\maketitle
\thispagestyle{plain}

\begin{abstract}
Evaluating software engineering agents on realistic benchmarks is costly,
since each task may require multi-step code exploration, modification, and
test execution. Existing efficient evaluation methods select representative
subsets to estimate full-benchmark performance, but are largely
\emph{result-only}: they fit historical pass/fail response matrices or static
task semantics, discarding how agents solve problems. We propose \ours, a
Privileged Trajectory-Aware Item Response Theory framework that fuses process
and outcome signals. Historical
execution trajectories supply process-level evidence beyond pass/fail, such as
explored context, attempted edits, and solving paths, which \ours uses as
privileged information for calibration subset selection and ability estimation.
Under low calibration budgets, \ours consistently outperforms prior IRT
baselines on score and ranking recovery across four SWE benchmarks.
Code and data are publicly available at
\url{https://github.com/DeepSoftwareAnalytics/PTA-IRT}.
\end{abstract}

\section{Introduction}

As Large Language Models (LLMs) are increasingly deployed as software
engineering agents, more challenging benchmarks emerge to evaluate their
ability to solve real-world software engineering tasks.
However, these agentic benchmarks make evaluation substantially more expensive
than conventional single-turn settings.
Repository-level benchmarks require agents to reason over
long horizons, explore codebases, invoke tools, edit source files, and execute
tests.
These multi-step interactions result in substantial inference costs and long
evaluation cycles.
For example, SWE-bench contains more than 2{,}000 programming tasks; a single
full-benchmark SWE-agent run has an estimated upper-bound cost exceeding
\$8{,}000 under a \$4-per-task limit, while the average cost on successfully
resolved instances is \$1.59 for SWE-agent with GPT-4 Turbo~\cite{yang2024swe,kapoor2025agents}.
This motivates the central question of our work: can we evaluate a new agent
on only a small budgeted subset of tasks while reliably recovering its
performance and relative ranking on the full benchmark?

To address this question, prior work on efficient benchmarking introduces
Item Response Theory (IRT) to select a small but representative subset of
benchmark instances and use limited observations to estimate full-benchmark
performance~\cite{tsutsumi2021deep,polo2024tinybenchmarks,kipnis2025metabench,zhou2026lost}. 
IRT-based methods fit historical response matrices to estimate item measurement characteristics
and to infer a model's latent ability from a limited set of observed responses. 
IRT therefore provides a promising framework for budgeted SWE agent evaluation.
However, prior IRT methods for efficient benchmarking typically represent each
task solely through its final pass/fail response, while a SWE agent's outcome is
produced through a multi-step trajectory of reasoning, context exploration,
tool use, code editing, and verification. Reducing a trajectory to a binary
outcome discards crucial process-level measurement signals about how an agent
reaches success or failure, including what context it explores, which actions
or edits it attempts, and where its problem-solving process breaks down.

To address this limitation, we propose \textbf{\ours}
(\textbf{P}rivileged \textbf{T}rajectory-\textbf{A}ware \textbf{I}tem
\textbf{R}esponse \textbf{T}heory), which incorporates historical agent
trajectories as privileged information into IRT-based data selection and
evaluation. \ours first converts trajectories into structured semantic
summaries and uses them to learn trajectory-aware four-parameter logistic (4PL) item measurement
characteristics. It then selects a difficulty-stratified calibration subset
using trajectory-aware Fisher information and transfers measurement knowledge to 
an ability estimator through Learning Using Privileged Information (LUPI). 
Given a new agent, \ours executes only the selected calibration
tasks and uses limited calibration observations to estimate its full-benchmark
performance and ranking.

\begin{table*}[t]
\centering
\small
\setlength{\tabcolsep}{3.2pt}
\begin{tabular}{lcccccc}
\toprule
 & \multicolumn{2}{c}{Dataset curation} & \multicolumn{4}{c}{Budgeted evaluation (subset$\rightarrow$full)} \\
\cmidrule(lr){2-3} \cmidrule(lr){4-7}
Dimension & Lite & Verified & Classical IRT & Deep/PSN-IRT & AutoJudger & \textbf{\ours} \\
\midrule
Target domain
  & SWE
  & SWE
  & General
  & General
  & Multimodal
  & SWE \\
Observation signal
  & ---
  & ---
  & Pass/Fail
  & Pass/Fail
  & Pass/Fail
  & Pass/Fail$+$Agent Traj. \\
Modeling method
  & Heuristic rules
  & Human judgment
  & Parametric IRT
  & Neural IRT
  & IRT $+$ LLM selection
  & Traj.-aware IRT \\
LLM involvement
  & ---
  & ---
  & ---
  & ---
  & Selects tasks
  & Summarizes traj. \\
Human involvement
  & ---
  & Heavy
  & ---
  & ---
  & ---
  & --- \\
\bottomrule
\end{tabular}
\caption{Comparison of efficient benchmarking for SWE evaluation.
\emph{Dataset curation}: SWE-bench Lite and Verified are fixed smaller subsets filtered from SWE-bench (Lite: rule-based; Verified: human annotation).
\emph{Budgeted evaluation} recovers full-benchmark scores from a calibration subset.
``---'' denotes not applicable / not involved.}
\label{tab:motivation}
\end{table*}

To validate the effectiveness of \ours, we conduct experiments on four
SWE benchmarks and compare it with prior IRT baselines.
Our results show that:
\begin{itemize}
    \item Under low calibration budgets, \ours consistently outperforms prior
    IRT baselines on score and ranking recovery.
    \item Privileged trajectory information is the source of these gains:
    outcome-only modeling or corrupted trajectory summaries weaken recovery.
    \item Structured trajectory summaries carry process signals beyond issue
    text and submitted patches, making budgeted evaluation more informative.
\end{itemize}

Our contributions are as follows:
\begin{itemize}
    \item We formulate efficient SWE benchmarking as a budgeted
    subset-to-full evaluation problem under a realistic protocol in which
    a new agent is executed only on the calibration subset, while historical
    trajectories are used offline for training.
    \item We introduce \ours, which incorporates historical agent trajectories
    as privileged information into IRT-based item selection and full-benchmark
    performance estimation.
    \item Extensive experiments on four SWE benchmarks show that \ours improves budgeted score and ranking recovery over prior IRT baselines.
\end{itemize}

\section{Background}

\begin{figure*}[t]
    \centering
    \includegraphics[width=\textwidth]{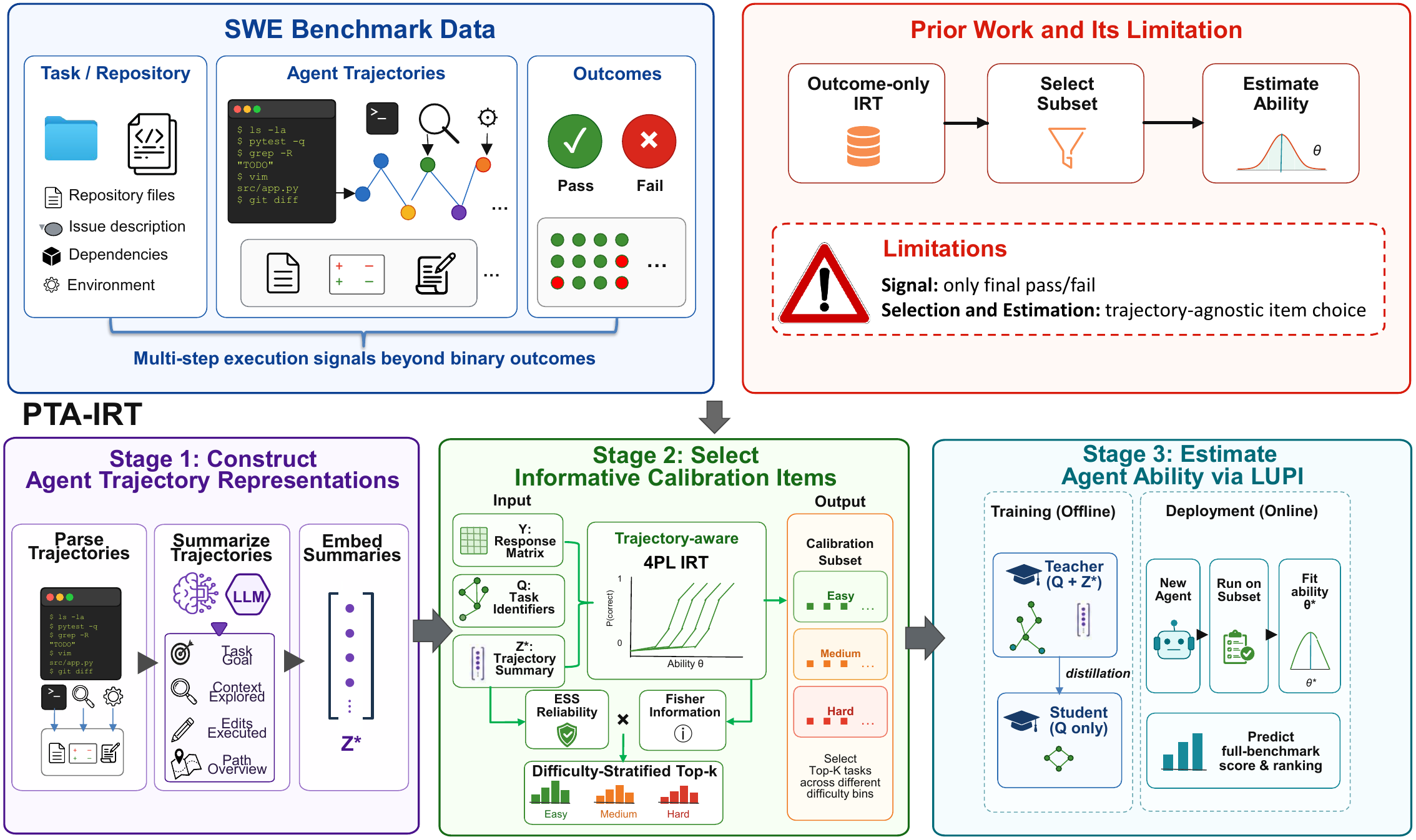}
    \caption{Limitations of prior work and an overview of \ours.}
    \label{fig:pipeline}
\end{figure*}

\subsection{Related Work}

Large language models and LLM-based agents have reshaped a wide range of
software engineering tasks~\cite{wang2025agents,liu2024large,wang2026context}.
On code generation, they now cover function-level synthesis as well as
repository-level completion, secure generation, and token-efficient
decoding~\cite{chen2021evaluatinglargelanguagemodels,jain2025livecodebench,zheng2024top,zheng2025humanevo,li2024repomincoder,liu2026shortcoderknowledgeaugmentedsyntaxoptimization,zhang2025llm,wang2026arkrepobench,wang2026realsec}.
Code search and retrieval-augmented generation further connect
natural-language intent to relevant code context~\cite{gong2025cosqa,wang2023you,gu2025retrieve,jiang2025aligncoder,wang2025draincode}.
Code summarization and comment generation recover intent from source
context~\cite{wang2025context,guo2023snippet}, while retrieval-augmented
commit message generation links edits to natural-language
descriptions~\cite{shi2022race}.
Issue resolution has become a central testbed for agentic software
engineering, with systems that explore repositories, apply patches, and
verify fixes against tests~\cite{jimenez2024swe,yang2024swe,tao2024magis,guo2025omnigirl,guo2026swe,jiang2026phoenixrepairrethinkingrepairstrategy,ye2025adverintent,zheng2026sweprimefewertrajectoriesbetter}.
Related directions include debugging, repository-level reasoning, code
translation, and code review~\cite{tian2024debugbench,wang2026reporeasoner,wang2025repotransbench,zheng2026staticdynamicbenchmarkingrealworld}.

These advances have driven more demanding repository-level and agentic
benchmarks, which require long-horizon reasoning, tool use, editing, and test
execution~\cite{jimenez2024swe,yang2025swe,deng2025swebenchproaiagents,guo2026swe,guo2025omnigirl,wang2026arkrepobench,wang2026realsec,zheng2024top,wang2025repotransbench,zheng2025humanevo,zheng2026staticdynamicbenchmarkingrealworld}.
Such evaluations are substantially more expensive than conventional
single-turn settings~\cite{chen2021evaluatinglargelanguagemodels,jain2025livecodebench}.

Recent research therefore examines the efficiency and reliability
of LLM benchmarks, with a growing interest in reducing evaluation cost
through a smaller set of test instances~\cite{perlitz2024efficient,vivek2024anchor,yuan2025beyondonesizefitsall}. IRT
provides a data-driven, item-level framework for this purpose~\cite{lord2008statistical}. By modeling
agents' responses across instances, IRT estimates item properties, such as
difficulty and discriminability, together with the latent abilities of
evaluated models.

In natural language processing, IRT was initially used to diagnose
fundamental properties of datasets and test items, including item
difficulty and discriminability~\cite{lalor2016building}. More recently,
its principles have been applied to efficient evaluation.
Deep-IRT and PSN-IRT extend conventional
IRT with neural architectures and more flexible response functions to model
complex model--item interactions~\cite{tsutsumi2021deep,zhou2026lost}. In a
different direction, AutoJudger proposes an agent-driven framework for
efficient benchmarking of multimodal large language models
\cite{ding2026autojudger}. 

However, IRT remains underexplored for SWE agent evaluation specifically,
where agent--instance response data are sparse and trajectories are long.
Instead, current SWE benchmarks still rely on rule-based filtering or human
verification to reduce evaluation cost, as in the curated SWE-bench Lite and
Verified subsets~\cite{jimenez2024swe}.
Table~\ref{tab:motivation} contrasts these approaches along target domain,
observation signal, modeling method, and human/LLM involvement.

\subsection{Preliminary: Item Response Theory}

IRT models the response behavior between
respondents and items through respondents' latent abilities and items'
measurement properties~\cite{hambleton2013item,reeve2005applying}. In SWE agent evaluation, agents correspond to
respondents, while benchmark instances correspond to items.

IRT uses an Item Characteristic Curve (ICC) to model the conditional
probability of a correct response. Under the one-parameter logistic model,
the probability that agent $i$ successfully solves instance $j$
is defined as
\begin{equation}
    P(X_{ij}=1 \mid \theta_i)
    =
    \frac{1}{1+\exp[-(\theta_i-b_j)]},
    \label{eq:1pl-irt}
\end{equation}
where $\theta_i$ denotes the latent ability of agent $i$
and $b_j$ denotes the difficulty parameter of instance $j$.

As IRT research advances, subsequent work incorporates additional item
parameters, explores more expressive ICC functions, and adopts deep-learning
approaches to improve its modeling capacity in complex evaluation 
settings~\cite{tsutsumi2021deep,zhou2026lost}.

\subsection{Preliminary: Learning Using Privileged Information}

LUPI leverages additional information
that is available during training but unavailable at test time
\cite{vapnik2009new,sharmanska2013learning}. Formally, each training example is
represented as $(\mathbf{x}, \mathbf{x}^{*}, y)$, where $\mathbf{x}$ denotes
the standard input, $\mathbf{x}^{*}$ denotes privileged information, and $y$
is the supervision signal. The goal of LUPI is to use
$\mathbf{x}^{*}$ during training to improve a predictor that relies only on
$\mathbf{x}$ at inference time.

A neural realization of LUPI uses a teacher--student paradigm, in which the
teacher accesses privileged information and transfers its knowledge to a
student through distillation. The student then relies only on the standard
input, without requiring privileged information at inference time
\cite{vapnik2015similarity,lopez2016unifying,lee2020learning}.

In our setting, agent trajectories serve as privileged
information, as they provide process-level evidence beyond binary responses.
They are available for historical agent--item interactions, but are not
required when estimating a new agent's performance on the full benchmark from
its responses on a calibration subset.

\section{\ours}

\subsection{Overview}

Evaluating a SWE agent on a full benchmark is costly. We study how to construct a small calibration subset from a full benchmark under a limited evaluation budget, and how to accurately estimate an agent's full-benchmark performance using only its outcomes on this subset.
As illustrated in Figure~\ref{fig:pipeline}, \ours consists of three consecutive stages:
\begin{itemize}
    \item \textbf{Stage~1: Construct Agent Trajectory Representations.}
     We compress heterogeneous and lengthy execution logs into comparable structured process summaries,
     preserving process-level evidence that binary pass/fail labels cannot capture.

    \item \textbf{Stage~2: Select Informative Calibration Items.}
    We fit a trajectory-aware 4PL IRT model, score tasks by Fisher information weighted by trajectory reliability,
    and select a difficulty-stratified calibration subset that is both informative and representative.

    \item \textbf{Stage~3: Estimate Agent Ability via LUPI.}
    On the same trajectory-aware 4PL backbone, a teacher--student objective transfers privileged process knowledge to a deployable student.
     At test time, we fit only a scalar ability on the calibration subset and extrapolate to the full benchmark.
\end{itemize}

\subsection{Construct Agent Trajectory Representations}

Raw agent trajectories contain lengthy commands, tool outputs, and intermediate
messages that vary across agents and frameworks.
We therefore build a trajectory parser to extract action steps from
heterogeneous logs, and design a prompt protocol that compresses each
historical trajectory into a structured \emph{process summary}.

The summary has four fields: \emph{Task Goal}, \emph{Context Explored},
\emph{Edits Executed}, and \emph{Path Overview}. 
The summarization model receives the task description, extracted action steps,
the benchmark reference patch, and the agent's submitted patch.
The reference patch serves to align task-relevant files, APIs, and change
scope across heterogeneous trajectories, and is not treated as an outcome
label.
The protocol further instructs the summarizer to describe observable actions
and edits, without inferring correctness or unobserved reasoning.
This compression preserves process-level evidence absent from binary
pass/fail labels.

We encode process summaries as $Z=\{\mathbf{z}_{ij}\}$ and use them together
with the response matrix $Y=\{y_{ij}\}$ during offline training.
Under our protocol, a new agent is evaluated on the calibration subset to
estimate full-benchmark performance, so process summaries are available for
historical interactions during training but unavailable for unevaluated tasks
at test time.

\subsection{Select Informative Calibration Items}

This stage constructs a calibration subset $\mathcal{S}$ that reduces uncertainty about agent ability while remaining representative of full-benchmark difficulty. Purely maximizing item information tends to select overly hard, mutually similar tasks and harms score recovery. We therefore select items by a trajectory-aware information score under difficulty stratification.

\paragraph{Trajectory-aware scorer.}
We fit a 4PL IRT model on historical agents $\mathcal{A}$ only. Each agent $i$ has latent ability $\theta_i$, and each task $j$ has global parameters $(a_j,b_j,c_j,d_j)$ for discriminability, difficulty, guessing, and feasibility ceiling. Under these parameters, the response probability is
\begin{equation}
    p_{ij}=c_j+(d_j-c_j)\,\sigma\!\bigl(a_j(\theta_i-b_j)\bigr),
\end{equation}
with $0\leq c_j<d_j\leq1$. Architecturally, an agent encoder maps a one-hot agent identifier to $\theta_i$, an item encoder maps a one-hot item identifier to $(a_j,b_j,c_j,d_j)$, and a summary encoder maps each privileged summary $\mathbf{z}_{ij}$ to bounded residuals $(\Delta a_{ij},\Delta b_{ij})=\alpha\tanh(g(\mathbf{z}_{ij}))$. Each encoder is a two-layer multilayer perceptron (MLP). On cells with a usable summary ($m_{ij}=1$), we form effective parameters
\begin{equation}
\begin{aligned}
    a_{ij} &= a_j + m_{ij}\Delta a_{ij},
    \quad
    b_{ij} = b_j + m_{ij}\Delta b_{ij}, \\
    p_{ij} &= c_j+(d_j-c_j)\,\sigma\!\bigl(a_{ij}(\theta_i-b_{ij})\bigr).
\end{aligned}
\end{equation}
Here $(a_j,b_j,c_j,d_j)$ remain shared item parameters for the agent population. A process summary provides agent--item interaction evidence, which we inject as residuals on discriminability and difficulty, i.e., the parameters that govern the ability-linked transition of the item characteristic curve. In contrast, the asymptotes $c_j$ and $d_j$ encode structural bounds of the task and evaluation setting, namely chance success and feasibility limits. These bounds are properties of the item and harness rather than of a single solving path, so we do not make them trajectory-dependent.

\paragraph{Information-aware stratified selection.}
We score each task by how much its outcomes inform ability estimation, then select under difficulty-stratified budget allocation. The contribution of interaction $(i,j)$ is the 4PL Fisher information~\cite{lord1980applications}
\begin{equation}
    I_{ij}(\theta_i)=
    \frac{a_{ij}^{2}\big(p_{ij}-c_j\big)^{2}\big(d_j-p_{ij}\big)^{2}}
    {(d_j-c_j)^{2}p_{ij}(1-p_{ij})}.
\end{equation}
Aggregating over $\mathcal{A}$ with offline quality weights $\omega_{ij}$ yields
\begin{equation}
\begin{array}{lcl}
  \operatorname{Fisher}(j) & = &
  \dfrac{\sum_{i\in\mathcal{A}}\omega_{ij}I_{ij}(\theta_i)}{\sum_{i\in\mathcal{A}}\omega_{ij}},
  \\[6pt]
  \operatorname{ESS}(j) & = &
  \displaystyle\sum_{i\in\mathcal{A}}m_{ij}\,\omega_{ij},
  \\[6pt]
  \operatorname{Info}(j) & = &
  \operatorname{Fisher}(j)\cdot\log\bigl(1+\operatorname{ESS}(j)\bigr).
\end{array}
\end{equation}
Here $m_{ij}\in\{0,1\}$ marks a usable process summary, and $\omega_{ij}\in[0,1]$ down-weights incomplete or malformed trajectory summaries.
$\operatorname{ESS}(j)$ is the effective sample size of privileged evidence on item $j$.
Multiplying $\operatorname{Fisher}(j)$ by $\log(1+\operatorname{ESS}(j))$ therefore reduces the score of items whose privileged evidence is of low quality.

Let $r_j=\frac{1}{|\mathcal{A}|}\sum_{i\in\mathcal{A}}y_{ij}$ be the pass rate of task $j$ on historical agents only. We partition tasks into difficulty bins by $r_j$, allocate the budget $k$ across bins in proportion to bin size, and within each bin select the tasks with the largest $\operatorname{Info}(j)$. The resulting subset $\mathcal{S}$ therefore favors high-information items while covering easy-to-hard regimes, which is essential for recovering full-benchmark scores from a small calibration budget.

\subsection{Estimate Agent Ability via LUPI}

A new agent is evaluated only on the calibration subset $\mathcal{S}$, and process summaries arise only after the agent is actually executed. Full-benchmark summaries are therefore unavailable at scoring time, which matches the LUPI setting. We design a teacher--student estimator on the same trajectory-aware 4PL backbone: privileged summaries supervise the teacher offline, while the student uses only identifiers and shared item parameters, estimates ability from $\mathcal{S}$, and extrapolates to all tasks.

\paragraph{Teacher--student training.}
The teacher and student share the agent, item, and summary encoders. The student predicts with global item parameters $(a_j,b_j,c_j,d_j)$. The teacher predicts with trajectory-adjusted parameters $(a_{ij},b_{ij})$, so the training objective depends on process-summary information. We minimize
\begin{equation}
\begin{aligned}
\mathcal{L}
&=
\operatorname{BCE}(p^{\mathrm{S}}_{ij},y_{ij})
\\
&\quad
+\,\lambda_{\mathrm{t}}\,m_{ij}\omega_{ij}\operatorname{BCE}(p^{\mathrm{T}}_{ij},y_{ij})
\\
&\quad
+\,\lambda_{\mathrm{KL}}\,m_{ij}\omega_{ij}
\operatorname{KL}\!\bigl(p^{\mathrm{T}}_{ij}\,\|\,p^{\mathrm{S}}_{ij}\bigr)
\\
&\quad
+\,\lambda_{\Delta}\,m_{ij}\bigl(\|\Delta a_{ij}\|^{2}+\|\Delta b_{ij}\|^{2}\bigr),
\end{aligned}
\end{equation}
where $p^{\mathrm{S}}_{ij}$ and $p^{\mathrm{T}}_{ij}$ are the student and teacher pass probabilities. The teacher and KL terms distill summary-conditioned measurements into the student. The residual penalty keeps $(\Delta a,\Delta b)$ small.

\paragraph{Test-time estimation.}
We freeze all network weights and fit a scalar ability $\theta_*$ on $\mathcal{S}$ by L-BFGS~\cite{Liu1989bfgs} from the observed outcomes, optionally with teacher distillation on $\mathcal{S}$. With $\theta_*$ and the shared item parameters $(a_j,b_j,c_j,d_j)$, the student predicts pass probabilities on every benchmark task and aggregates them into overall performance and ranking.

\section{Experimental Setup}
\subsection{Datasets}
As shown in Table~\ref{tab:benchmark-overview}, we evaluate on four SWE-bench versions (Lite, Verified, Full, and Pro).
Lite and Verified are curated subsets of Full~\cite{jimenez2024swe}: Lite uses rule-based filtering toward more self-contained functional fixes, while Verified uses human annotation to remove underspecified or problematic instances.
SWE-bench Pro~\cite{deng2025swebenchproaiagents} further extends SWE-bench to long-horizon, multi-file enterprise-style tasks across broader repositories and languages.
Because Lite and Verified are the community's default reporting targets, they attract substantially more public agent submissions with parseable trajectories than Full or Pro.
We therefore obtain denser model coverage on Lite and Verified, while Full and Pro emphasize larger task collections under sparser agent pools.
We use each official release and retain all models with complete, successfully parsed trajectories.

\begin{table}[t]
    \centering
    \begin{tabular}{lrr}
    \toprule
    \textbf{Benchmark} & \textbf{\# Tasks} & \textbf{\# Evaluated Models} \\
    \midrule
    SWE-bench Lite        & 300   & 35 \\
    SWE-bench Verified    & 500   & 70 \\
    SWE-bench Full        & 2,294 & 14 \\
    SWE-bench Pro         & 730   & 14 \\
    \bottomrule
    \end{tabular}
    \caption{Overview of the benchmarks and evaluated models.}
    \label{tab:benchmark-overview}
\end{table}

\begin{table*}[t]
\centering
\small
\setlength{\tabcolsep}{2.4pt}
\begin{tabular}{l*{15}{c}}
\toprule
 & \multicolumn{3}{c}{Lite} & \multicolumn{3}{c}{Verified} & \multicolumn{3}{c}{Full} & \multicolumn{3}{c}{Pro} & \multicolumn{3}{c}{Avg} \\
\cmidrule(lr){2-4} \cmidrule(lr){5-7} \cmidrule(lr){8-10} \cmidrule(lr){11-13} \cmidrule(lr){14-16}
Method & MAE$\downarrow$ & $\tau$$\uparrow$ & $\rho$$\uparrow$ & MAE$\downarrow$ & $\tau$$\uparrow$ & $\rho$$\uparrow$ & MAE$\downarrow$ & $\tau$$\uparrow$ & $\rho$$\uparrow$ & MAE$\downarrow$ & $\tau$$\uparrow$ & $\rho$$\uparrow$ & MAE$\downarrow$ & $\tau$$\uparrow$ & $\rho$$\uparrow$ \\
\midrule
MLE & .164$\pm$.022 & .718 & .894 & .235$\pm$.045 & .648 & .831 & .120$\pm$.056 & .319 & .424 & .165$\pm$.039 & .429 & .675 & .171$\pm$.041 & .528 & .706 \\
MCMC & .217$\pm$.048 & .647 & .805 & .130$\pm$.041 & .769 & .923 & .073$\pm$.050 & .780 & .921 & .084$\pm$.041 & .846 & .952 & .126$\pm$.045 & .761 & .900 \\
VI & .239$\pm$.027 & .731 & .898 & .142$\pm$.012 & .855 & .967 & .281$\pm$.140 & .934 & .982 & .226$\pm$.028 & .868 & .952 & .222$\pm$.052 & .847 & .950 \\
VIBO & .206$\pm$.031 & .691 & .866 & .291$\pm$.036 & .583 & .737 & .157$\pm$.069 & .890 & .960 & .211$\pm$.057 & .692 & .864 & .216$\pm$.048 & .714 & .857 \\
Deep-IRT & .305$\pm$.145 & .682 & .863 & .191$\pm$.049 & .724 & .909 & .362$\pm$.120 & .934 & .982 & .289$\pm$.144 & .692 & .868 & .287$\pm$.114 & .758 & .905 \\
PSN-IRT & .114$\pm$.035 & .650 & .817 & .216$\pm$.045 & .786 & .941 & .198$\pm$.133 & .758 & .921 & .073$\pm$.013 & .692 & .886 & .150$\pm$.057 & .722 & .891 \\
AutoJudger & .128$\pm$.014 & .620 & .826 & .200$\pm$.011 & .673 & .871 & .151$\pm$.051 & .626 & .736 & .120$\pm$.047 & .626 & .807 & .150$\pm$.031 & .637 & .810 \\
\midrule
\textbf{PTA-IRT} & \textbf{.045$\pm$.004} & \textbf{.836} & \textbf{.950} & \textbf{.043$\pm$.015} & \textbf{.872} & \textbf{.976} & \textbf{.048$\pm$.025} & \textbf{.956} & \textbf{.991} & \textbf{.029$\pm$.017} & \textbf{.890} & \textbf{.974} & \textbf{.041$\pm$.015} & \textbf{.888} & \textbf{.973} \\
\bottomrule
\end{tabular}
\caption{Comparison at a $10\%$ calibration budget across four SWE benchmarks. Lower MAE and higher Kendall's $\tau$ / Spearman's $\rho$ are better. Best values in each column are bolded (all ties are bolded). Values are rounded to three decimals with leading zeros omitted.}
\label{tab:rq1}
\end{table*}

\begin{figure*}[t]
    \centering
    \includegraphics[width=0.9\textwidth]{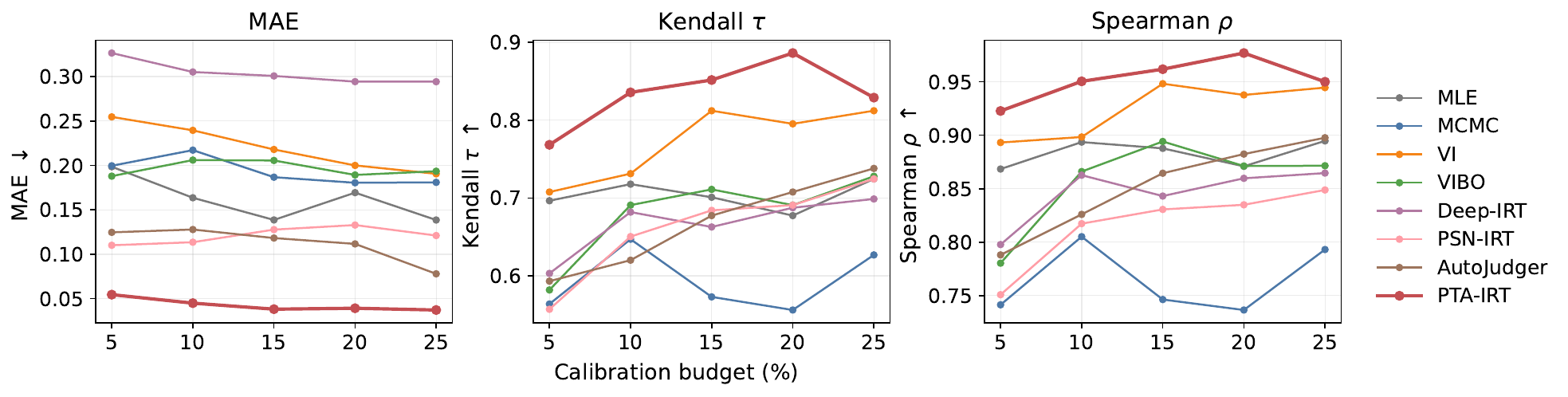}
    \caption{Budget sensitivity on SWE-bench Lite under $5$--$25\%$ calibration.
    Each panel shows one metric (MAE, Kendall's $\tau$, and Spearman's $\rho$).}
    \label{fig:budget-lite}
\end{figure*}

\subsection{Baselines}

We compare \ours against a representative set of prior IRT baselines spanning classical IRT, neural IRT, and agent-driven selection.
Classical baselines include Maximum Likelihood Estimation (MLE), Markov Chain Monte Carlo (MCMC)~\cite{hastings1970monte}, Variational Inference (VI)~\cite{jordan1999introduction}, and VIBO~\cite{wu2020variational}.
Advanced methods include Deep-IRT~\cite{tsutsumi2021deep} and PSN-IRT~\cite{zhou2026lost}.
We additionally include AutoJudger~\cite{ding2026autojudger}, which combines IRT-based difficulty estimation with an LLM agent that adaptively selects calibration items.

\subsection{Metrics}

We adopt three metrics: Mean Absolute Error (MAE), Kendall's $\tau$, and Spearman's $\rho$.

\begin{itemize}
    \item \textbf{MAE:}
    the average deviation between predicted and ground-truth full-benchmark performance,
    $\mathrm{MAE} = \frac{1}{N}\sum_{i=1}^{N} \left| \hat{y}_i - y_i \right|$,
    where $\hat{y}_i$ and $y_i$ are the predicted and ground-truth scores of model $i$, and $N$ is the number of evaluated models.

    \item \textbf{Kendall's $\tau$:}
    pairwise ordinal agreement between the predicted and ground-truth rankings,
    $\tau = \frac{C-D}{\binom{N}{2}}$,
    where $C$ and $D$ are the numbers of concordant and discordant model pairs.

    \item \textbf{Spearman's $\rho$:}
    the global monotonic correlation between the predicted and ground-truth rankings,
    $\rho = \operatorname{corr}\bigl(\operatorname{rank}(\hat{\mathbf{y}}), \operatorname{rank}(\mathbf{y})\bigr)$,
    where $\hat{\mathbf{y}}$ and $\mathbf{y}$ are the vectors of predicted and ground-truth performance.
\end{itemize}

\subsection{Implementation Details}
To ensure reliability, we use four-fold cross-validation: in each fold, 75\% of the models are used for training and the remaining 25\% for testing.
Process summaries are generated with DeepSeek-V4-Flash\footnote{\raggedright\url{https://huggingface.co/deepseek-ai/DeepSeek-V4-Flash}} and embedded with all-MiniLM-L6-v2.\footnote{\raggedright\url{https://huggingface.co/sentence-transformers/all-MiniLM-L6-v2}}

\section{Main Results}

\subsection{RQ1: Effectiveness of \ours}

\begin{table*}[t]
\centering
\small
\setlength{\tabcolsep}{1.8pt}
\begin{tabular}{l*{15}{c}}
\toprule
 & \multicolumn{3}{c}{Lite} & \multicolumn{3}{c}{Verified} & \multicolumn{3}{c}{Full} & \multicolumn{3}{c}{Pro} & \multicolumn{3}{c}{Avg} \\
\cmidrule(lr){2-4} \cmidrule(lr){5-7} \cmidrule(lr){8-10} \cmidrule(lr){11-13} \cmidrule(lr){14-16}
Method & MAE$\downarrow$ & $\tau$$\uparrow$ & $\rho$$\uparrow$ & MAE$\downarrow$ & $\tau$$\uparrow$ & $\rho$$\uparrow$ & MAE$\downarrow$ & $\tau$$\uparrow$ & $\rho$$\uparrow$ & MAE$\downarrow$ & $\tau$$\uparrow$ & $\rho$$\uparrow$ & MAE$\downarrow$ & $\tau$$\uparrow$ & $\rho$$\uparrow$ \\
\midrule
\textbf{PTA-IRT} & \textbf{.045$\pm$.004} & \textbf{.836} & \textbf{.950} & .043$\pm$.015 & \textbf{.872} & \textbf{.976} & .048$\pm$.025 & \textbf{.956} & \textbf{.991} & \textbf{.029$\pm$.017} & \textbf{.890} & \textbf{.974} & \textbf{.041$\pm$.015} & \textbf{.888} & \textbf{.973} \\
w/o Traj.\ scorer & .055$\pm$.021 & .778 & .930 & .065$\pm$.008 & .860 & .969 & \textbf{.048$\pm$.029} & .912 & .982 & .047$\pm$.010 & \textbf{.890} & .969 & .054$\pm$.017 & .860 & .963 \\
w/o LUPI & .085$\pm$.004 & .573 & .755 & \textbf{.042$\pm$.013} & .848 & .964 & .053$\pm$.032 & .802 & .912 & .092$\pm$.040 & .868 & .965 & .068$\pm$.022 & .773 & .899 \\
\midrule
+ Top-K & .196$\pm$.031 & .812 & .942 & .193$\pm$.012 & .752 & .917 & .228$\pm$.163 & .934 & .982 & .086$\pm$.025 & .802 & .921 & .176$\pm$.058 & .825 & .941 \\
+ Clustering & .182$\pm$.037 & .819 & .947 & .189$\pm$.015 & .757 & .921 & .225$\pm$.162 & .912 & .974 & .090$\pm$.027 & .758 & .908 & .171$\pm$.060 & .812 & .937 \\
\bottomrule
\end{tabular}
\caption{Ablation of PTA-IRT at a $10\%$ calibration budget. Lower MAE and higher Kendall's $\tau$ / Spearman's $\rho$ are better. Best values in each column are bolded (all ties are bolded). Values are rounded to three decimals with leading zeros omitted.}
\label{tab:ablation}
\end{table*}

\begin{figure*}[t]
    \centering
    \includegraphics[width=\textwidth]{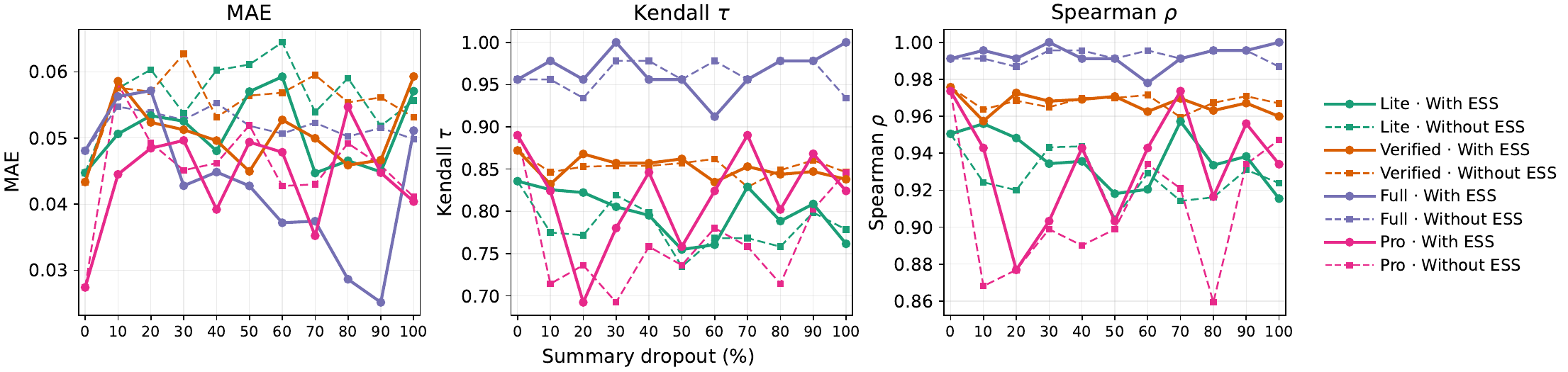}
    \caption{Contribution of trajectory summaries under summary dropout ($10\%$ calibration).
    Solid lines: With ESS; dashed lines: Without ESS. Colors denote benchmarks.}
    \label{fig:summarydropout}
\end{figure*}

RQ1 examines how \ours performs under limited calibration budgets relative to prior IRT baselines.
Table~\ref{tab:rq1} reports the main comparison at a $10\%$ budget on four SWE benchmarks, and Figure~\ref{fig:budget-lite} evaluates budget sensitivity on SWE-bench Lite under $5$--$25\%$ calibration.

\noindent\textbf{Finding 1.}
\textit{With only $10\%$ calibration, \ours consistently outperforms prior IRT baselines on both score recovery and ranking.}
As shown in Table~\ref{tab:rq1}, \ours is best on every MAE/$\tau$/$\rho$ column across the four SWE benchmarks, averaging MAE $0.041\!\pm\!0.015$, $\tau\!=\!0.888$, and $\rho\!=\!0.973$.
This advantage holds under two complementary regimes.
On Lite and Verified, where agent pools are dense, \ours attains the lowest MAE and the highest $\tau$/$\rho$.
On Full and Pro, where task sets are large but agents are few, \ours likewise achieves the best score and ranking recovery.

\noindent\textbf{Finding 2.}
\textit{\ours is robust across calibration budgets, and a small budget already yields practically useful ranking agreement.}
As shown in Figure~\ref{fig:budget-lite}, $5\%$ calibration already gives $\tau\!=\!0.768$, within the practical range reported in prior analyses~\cite{liu2021doqa,sun2023validity} ($\tau\!\geq\!0.7$--$0.8$).
Raising the budget from $5\%$ to $20\%$ further improves ranking on Lite ($\tau$: $0.768\!\rightarrow\!0.886$) and tends to lower MAE.
\ours remains the best method across the full $5$--$25\%$ range, underscoring the robustness of our approach under varying calibration budgets.

\subsection{RQ2: Ablation Analysis of \ours}

RQ2 examines the contribution of each component of \ours to recovering full-benchmark scores and rankings.
Table~\ref{tab:ablation} reports ablations at a $10\%$ calibration budget.
w/o Traj.\ scorer removes the summary encoder from the 4PL IRT used for informative selection.
w/o LUPI drops teacher--student LUPI and estimates ability with the student model alone, without privileged summary supervision.
+ Top-K and + Clustering replace stratified information selection with global Top-K by information and pass-rate clustering, respectively.

\noindent\textbf{Finding 3.}
\textit{Each component of \ours contributes to recovering full-benchmark scores and rankings.}
As shown in Table~\ref{tab:ablation}, the full configuration achieves the best average MAE/$\tau$/$\rho$.
Removing the trajectory-aware scorer (w/o Traj.\ scorer) or LUPI (w/o LUPI) raises Avg MAE and lowers ranking agreement; the only local exception is a marginal MAE improvement on Verified without LUPI, which does not overturn the average.
Replacing stratified selection with Top-K or clustering likewise worsens the average on both score and ranking.

\subsection{RQ3: Contribution of Trajectory Summaries}

RQ3 examines the contribution of historical trajectory summaries to \ours.
We apply \emph{summary dropout}: a controlled fraction of train-summary embeddings is set to zero before retraining the trajectory-aware scorer and rebuilding the stratified calibration set $\mathcal{S}$.
We compare With ESS, which down-weights corrupted cells, against Without ESS, which still feeds corrupted embeddings at full weight.
Figure~\ref{fig:summarydropout} reports MAE, Kendall's $\tau$, and Spearman's $\rho$ on four SWE benchmarks under a $10\%$ calibration budget.

\noindent\textbf{Finding 4.}
\textit{Trajectory summaries improve recovery quality; their benefit comes from summary content rather than a vacuous privileged channel.}
As shown in Figure~\ref{fig:summarydropout}, increasing summary dropout generally weakens score recovery and ranking relative to the clean setting, though the curves are not strictly monotonic across benchmarks.
At full dropout without ESS, Lite approaches the RQ2 ablation that removes the trajectory-aware scorer (MAE $0.056$, $\tau\!=\!0.778$), indicating that the gains of \ours mainly come from summary content rather than merely having an unused summary pathway.

\noindent\textbf{Finding 5.}
\textit{ESS effectively protects \ours when trajectory summaries are unreliable.}
Under the same dropout levels, With ESS stays closer to the clean setting than Without ESS on most benchmark--metric pairs.
The protective gap is clearest at low-to-mid dropout (e.g., on Lite, Without ESS already falls to $\tau\!\approx\!0.735$ at $50\%$ dropout), where ESS can still down-weight corrupted cells.
At extreme dropout, both degrade as usable process signal disappears.

\subsection{RQ4: Characteristics of Trajectory Summaries}

RQ4 examines what distinctive structure trajectory summaries capture relative to issue text and submitted patches, and how that structure is organized across summary fields and same-item outcome pairs.
We answer this with two complementary views: cross-channel similarity (Table~\ref{tab:rq4-sim}) and same-item outcome geometry (Table~\ref{tab:rq4-pair}).

\paragraph{Cross-channel similarity.}
For each (agent, item) cell we compute cosine similarity between the process summary (and its four fields) and (i)~the question embedding $\mathbf{q}_j$ and (ii)~the submitted-answer embedding.
As a control for question alignment, we also report $\Delta$ vs.\ randQ: the gap between similarity to the true question and the mean similarity to a randomly sampled question.
Table~\ref{tab:rq4-sim} summarizes the results.

\begin{table}[t]
\centering
\small
\setlength{\tabcolsep}{3.5pt}
\begin{tabular}{lrrr}
\toprule
\textbf{Channel} & $\leftrightarrow$Q & $\leftrightarrow$Ans. & $\Delta$ randQ \\
\midrule
Summary & 0.714 & 0.638 & $+0.473$ \\
\quad Task Goal & 0.776 & 0.532 & $+0.587$ \\
\quad Context Explored & 0.502 & 0.533 & $+0.314$ \\
\quad Edits Executed & 0.530 & 0.670 & $+0.333$ \\
\quad Path Overview & 0.468 & 0.441 & $+0.272$ \\
\bottomrule
\end{tabular}
\caption{Cross-channel cosine similarity on SWE-bench Verified.
$\Delta$ vs.\ randQ applies to the $\leftrightarrow$~Question column.}
\label{tab:rq4-sim}
\end{table}

\noindent\textbf{Finding 6.}
\textit{Trajectory summaries exhibit a clear division of labor: Task Goal tracks the issue text, Edits Executed tracks the submitted patch, while Context Explored and Path Overview remain process-specific.}
Whole-summary similarity to the question ($0.714$) exceeds a random-question baseline by a large margin ($\Delta\!=\!+0.473$).
Field-level scores reveal complementary roles rather than redundancy.
Task Goal is closest to the question ($0.776$), which largely explains the high Summary$\leftrightarrow$Question score.
Edits Executed is closest to the submitted answer ($0.670$), whereas Context Explored and Path Overview stay lower on both axes ($\leq\!0.533$).
A trajectory summary therefore combines question-facing goal understanding with edit-level content and independent exploration/path structure.

\paragraph{Same-item outcome geometry.}
We next compare representations of successful and failed solving paths on the \emph{same} task.
For every item with multiple agents, we form agent pairs and report mean embedding distance ($1-\mathrm{cosine}$) for both-pass (PP), both-fail (FF), and pass--fail (PF) pairs, together with $\Delta_{\mathrm{out}}=d(\mathrm{PF})-\tfrac12\bigl(d(\mathrm{PP})+d(\mathrm{FF})\bigr)$.
A larger $\Delta_{\mathrm{out}}$ indicates stronger outcome-related structure.
Table~\ref{tab:rq4-pair} reports Answer, Summary, and the four summary fields.

\begin{table}[t]
\centering
\small
\setlength{\tabcolsep}{3.5pt}
\begin{tabular}{lrrrr}
\toprule
\textbf{Channel} & PP & FF & PF & $\Delta_{\mathrm{out}}$ \\
\midrule
Answer & 0.175 & 0.290 & 0.265 & 0.033 \\
Summary & 0.143 & 0.163 & 0.158 & 0.005 \\
\quad Task Goal & 0.127 & 0.147 & 0.141 & 0.003 \\
\quad Context Explored & 0.337 & 0.387 & 0.372 & 0.010 \\
\quad Edits Executed & 0.366 & 0.422 & 0.414 & 0.020 \\
\quad Path Overview & 0.363 & 0.387 & 0.389 & 0.013 \\
\bottomrule
\end{tabular}
\caption{Same-item pair distances on SWE-bench Verified.
PP/FF/PF: both-pass / both-fail / pass--fail.}
\label{tab:rq4-pair}
\end{table}

\noindent\textbf{Finding 7.}
\textit{Submitted answers concentrate outcome geometry, while trajectory summaries preserve process diversity.}
For answers, same-outcome successes are much closer than failures (PP~$0.175$ vs.\ FF~$0.290$), and $\Delta_{\mathrm{out}}\!=\!0.033$.
For full summaries, PP and FF distances remain close ($0.143$ vs.\ $0.163$) and $\Delta_{\mathrm{out}}$ is only $0.005$; among fields, Context Explored, Edits Executed, and Path Overview keep large absolute distances ($\geq\!0.337$), and Edits Executed shows the largest $\Delta_{\mathrm{out}}$ ($0.020$), consistent with its stronger answer alignment in Table~\ref{tab:rq4-sim}.
Thus pass/fail is expressed most directly in the final patch, whereas summaries retain heterogeneous solving paths that are not collapsed by binary labels and patches alone.

\section{Conclusion}

Evaluating SWE agents on full benchmarks is expensive because each task can require long-horizon exploration, editing, and testing.
We present \ours, which treats historical execution trajectories as privileged information for efficient IRT-based evaluation.
\ours builds structured trajectory summaries, selects a difficulty-stratified calibration subset with trajectory-aware measurement features, and estimates a new agent's full-benchmark score and ranking from limited calibration observations via LUPI.
Under low calibration budgets, \ours consistently outperforms prior IRT baselines on score and ranking recovery across four SWE benchmarks under our evaluation protocol, showing that privileged trajectory information makes budgeted SWE agent evaluation more informative.

\bibliography{ref}

\end{document}